\documentclass[a4paper,twocolumn,11pt]{quantumarticle}
\pdfoutput=1

\usepackage[utf8]{inputenc}
\usepackage[english]{babel}
\usepackage[T1]{fontenc}

\usepackage{graphicx} % Required for inserting images
\usepackage{amsmath, amssymb, amsthm}
\usepackage{etoolbox}
\usepackage{quantikz}
\usepackage[numbers,sort&compress]{natbib}
\usepackage{verbatim}

\usepackage{hyperref}

\newtheorem{theorem}{Theorem}
\newtheorem{lemma}{Lemma}

\newcommand{\PR}{\mathbb{P}}
\newcommand{\EX}{\mathbb{E}}

\title{Optimal Resource Scaling for Early Fault-Tolerant Iterative Quantum Phase Estimation under Cost–Error Tradeoffs}

\author{Mrudula A Mahindrakar}
\affiliation{Department of Electrical Engineering,
Indian Institute of Technology Madras, India}
\email{ee25s011@smail.iitm.ac.in}
\author{Avhishek Chatterjee}
\affiliation{Department of Electrical Engineering,
Indian Institute of Technology Madras, India}
\email{avhishek@ee.iitm.ac.in}

\begin{document}

\maketitle

\begin{abstract}
    
The iterative quantum phase estimation algorithm (IPEA) is widely considered to be better suited for noisy intermediate-scale quantum (NISQ) and early fault-tolerant hardware compared to the original algorithm, which requires the inverse quantum Fourier transform. Current implementations of error mitigation and noise-adapted error correction lead to imperfect implementation of (powers of) the unitary and result in erroneous phase feedback in IPEA. A widely used statistical approach to improve the reliability of phase bits is to repeat the $k^{\text{th}}$ iteration for the $2^{k-1}$-th power of the unitary $N_k$ times (i.e. $N_k$ shots),  followed by a majority decision. However, due to variations in their circuit complexities and error mitigation techniques, different iterations incur varying resource costs (denoted $w_k$) and different per-shot error probabilities ($q_k$). 
Since NISQ and early fault-tolerant hardware are highly resource constrained, we ask what the optimal number of shots per bit (iteration) is:  
\begin{align*} & \arg \max_{\{N_k: 1 \le k \le L\}} \PR(\text{all } L \text{ bits correct}) \\ & ~~~~~~~~~~~~\text{ s.t.} \sum_{k=1}^L w_k N_k \le W.\end{align*}
We construct upper and lower surrogate objectives via concentration and anti-concentration inequalities and prove an asymptotic optimality guarantee for whenever the budget satisfies $W = \sum_{k=1}^L \frac{w_k}{c_k}(\ln L + \omega(1))$, where $c_k=-\ln\left(2\sqrt{\left(1-\frac{q_k}{2} \right) \frac{q_k}{2}}\right)$, which, for $q_k\ll 1$, is proportional to $\ln\frac{1}{q_k}$. Optimizing the tractable lower surrogate yields a true success probability within a $(1-o(1))$ factor of the exact optimum. We observe that when resources are abundant, i.e., $W \gg  \ln(L)\sum_{k=1}^L \frac{w_k}{c_k}$, optimal allocations are proportional to $\ 1/c_k$. However, when resources are sufficient but not abundant, i.e., $(\ln(L)+\omega(1))\sum_k \frac{w_k}{c_k}\le W \le O\left( \ln(L)\cdot \sum_k \frac{w_k}{c_k}\right)$, an additional $-\frac{1}{c_k}\ln\frac{w_k}{c_k}$ correction accounts for error propagation from less-significant to more-significant bits.
\end{abstract}

\section{Introduction}

Quantum phase estimation (QPE) is a fundamental primitive in quantum computing.  Phase estimation underlies a wide range of quantum algorithms, including order finding, 
Hamiltonian simulation, quantum chemistry algorithms, and eigenvalue estimation. In the standard quantum phase-estimation algorithm, 
a register of control qubits is prepared in a superposition, controlled powers $U^{2^{k-1}}$ are applied, and an inverse quantum Fourier transform is used to extract the 
phase~\cite{nielsenchuang}.

However, for noisy intermediate-scale quantum (NISQ) and early fault-tolerant (FT) quantum processors, this standard implementation presents a substantial resource challenge, particularly for the inverse Fourier transform circuits. For current hardware technologies, the iterative quantum phase estimation algorithm (IPEA) \cite{kitaev1995quantum,dobvicek2007}, in 
which the phase is inferred one bit at a time without the use of the inverse quantum Fourier transform, is an attractive alternative \cite{Koh2024IQPE_NISQ}.  

To determine successive phase bits of the unitary $U$, the IPEA algorithm requires experiments involving 
large powers of the unitary $U,\; U^2,\; U^4,\ldots$,
when the phase is an $ L$- bit number. The experiment involving $U^{2^{k-1}}$, known as the $k^\text{th}$ iteration of IPEA, recovers the $k^\text{th}$ bit of the $L$ bit phase.  

The resource cost for running one shot of iteration $k$, denoted by $w_k$, may vary with $k$, due to the hardware requirements. In this paper, $w_k$ captures the number of oracle calls, number of elementary gates, circuit depth, 
execution time, etc. in a single cost metric.  In NISQ and early fault-tolerant circuits, the resource cost for error mitigation or partial error correction is also part of $w_k$. This allows us to include scenarios where different iterations may employ different error mitigation or error correction circuits. Hence, for the purpose of this paper, the interpretation of $w_k$ is deliberately left general.

In a fully fault-tolerant circuit, the probability of obtaining an erroneous $k$-th bit using IPEA would be negligible. However, in NISQ or early FT era circuits, despite error mitigation and preliminary noise-adapted error corrections, this probability is non-negligible.  It depends on the circuit used in the $k^\text{th}$ iteration and also on the reliability of the outputs of iterations $k+1$ to $L$. The latter is due to the fact that for the $k^\text{th}$ iteration IPEA uses the bits obtained from $ k+1$ to $ L$ iterations as feedback.  

Due to insufficient error correction in NISQ and early FT hardware, even when $k+1$ to $L$ bits are reliably available, the probability of an erroneous measurement of the $k^\text{th}$ phase bit is non-negligible and depends on $k$ via circuit depth, number of oracle calls required for implementing $U^{2^k}$, choice of error mitigation technique, etc. Given $k+1$ to $L$ bits are reliably available, we denote the probability of obtaining the $k^\text{th}$ bit erroneously in a single execution (a.k.a. shot) of iteration k, by $\frac{q_k}{2}$. 

There is a simple physical interpretation for this.   Due to insufficient error correction in current hardware,  the pre-measurement quantum state in the $k^\text{th}$ iteration is a $q_k$-mixture of the desired state and an unknown state. This is equivalent to $q_k$-depolarizing noise acting on the output quantum state if that unknown state is completely unpredictable. In this case, the probability of obtaining the $k$-th phase bit incorrectly would be $\frac{q_k}{2}$.

Repeated execution of each iteration and classical aggregation, for example, through 
majority voting, can suppress statistical measurement error. The natural question is 
therefore not simply how many repetitions are sufficient at each iteration. Rather, the question is: how should a finite experimental budget be distributed among the different iterations to maximize the overall reliability of IPEA?

Importantly, the framework we consider in this paper does not assume that error mitigation is universally beneficial: a mitigation procedure that substantially reduces $\epsilon_k$ but incurs a sufficiently large increase in $w_k$ may provide little or no benefit under a fixed resource budget. The optimal repetition allocation consequently depends on the joint scaling of cost and error rather than on either quantity alone.

In this work, we formulate this question as a resource-constrained optimization problem for IPEA. Let $N_k$ denote the number of repetitions of the $k^\text{th}$ phase-estimation iteration. Given a total resource budget $W$, the admissible repetition schedules satisfy
\begin{equation}
    \sum_{k=1}^{L} w_k N_k \leq W. \nonumber
\end{equation}
Rather than prescribing a target error probability for each bit independently, we optimize 
the probability of recovering the complete $L$-bit phase, $ P_{\mathrm{succ}}(\{N_k\})
    =
    \Pr(\text{all }L\text{ phase bits are correct}).$
Our central problem is
\begin{align}
\max_{\{N_k\}}
\quad&
P_{\mathrm{succ}}(\{N_k\}) \nonumber
\\
\text{subject to}\quad&
\sum_{k=1}^{L} w_kN_k\leq W. \label{eq:main_optimization}
\end{align}

This formulation captures a resource-allocation problem that is intrinsic to iterative 
phase estimation: repetitions improve statistical reliability, whereas later iterations 
may simultaneously become more expensive and less reliable. To avoid algebraic issues, we assume $w_k \ge 1$ for all $k$, since we can always normalize the individual costs $\{w_k\}$ and the total budget $W$ with respect to $\min_{1\le k \le L} w_k$.

\subsection{Related Literature}

%%%%%%%%%%%%%%%%%%%%%%%%%%%%
Early work on quantum phase estimation established that Heisenberg-limited precision can be obtained by applying a phase-encoding operation at progressively increasing interrogation times. Higgins \textit{et al.} demonstrated this experimentally without entanglement and subsequently developed a nonadaptive protocol using predetermined interrogation depths \cite{Higgins2007,Higgins2009}. Kimmel \textit{et al.} built on this framework to introduce robust phase estimation (RPE), which estimates systematic gate errors while tolerating state-preparation and measurement errors \cite{kimmel2015}. Their RPE protocol is nonadaptive: measurements are performed repeatedly at progressively increasing sequence depths, and the number of repetitions at each depth is chosen to obtain robust estimation with Heisenberg scaling. Thus, although RPE involves a resource-allocation problem across different phase-amplification levels, it is distinct from the adaptive, bit-by-bit structure of IPEA.

Subsequent work has experimentally demonstrated RPE \cite{Rudinger2017}, investigated tests to detect violations of its robustness assumptions \cite{Russo2021}, and developed Bayesian post-processing to reduce its sampling overhead \cite{Hurant2024}. More recently, Dong \textit{et al.} introduced quantum signal-processing phase estimation as a low-depth alternative designed to improve robustness to realistic time-dependent errors \cite{Dong2025}.

However, these works do not consider the adaptive IPEA protocol studied here. On the contrary, they employ different phase estimation schemes, most notably nonadaptive RPE or quantum-signal-processing-based phase estimation. The error analysis of IPEA is significantly different due to the sequential, feedback-dependent iterations of IPEA. Consequently, their resource allocation and error analysis do not address the problem of optimally choosing the number of repetitions (shots) of different iterations of IPEA.

There is a line of work on IPEA that is closer to our work yet with significant differences. O'Loan introduced an iterative phase-estimation scheme and analyzed its performance in the presence of depolarizing noise, showing that estimation remains reliable below a noise-dependent limit on the number of iterations \cite{OLoan2010}. O'Malley \textit{et al.} experimentally implemented iterative phase estimation for molecular-energy estimation, using majority voting over repeated measurements for each phase bit \cite{OMalley2016}. Van den Berg subsequently optimized the sampling complexity of Kitaev's phase-estimation algorithm, in the absence of noise, using increasingly accurate phase shifts, reducing the number of measurements required for a prescribed accuracy \cite{VandenBerg2020}.

Higher noise levels cause irreversible feedback errors in IPEA. To mitigate that
 Bayesian phase estimation \cite{wiebe2016efficient,obrien2019quantum,dutkiewicz2022heisenberg} uses per-bit majority voting with  posterior update across circuit depth. However, this approach incurs significant classical inference overhead and circuit recompilation, which have to happen online. 
 
 In contrast, majority-voted IPEA admits static, offline shot schedules with negligible classical runtime cost. This motivates us to find an exact characterization of its optimal shot allocation under resource constraint when noise rates and resource cost for iterations are non-uniform across iterations, which is the case for most quantum simulations.

Allocating a constrained measurement budget across heterogeneous quantum circuits has been studied  for variational  estimation under shot noise and hardware decoherence \cite{arrasmith2020operator,kubler2020adaptive,kulkarni2026noise}. However, these works either distribute shots proportionally to Pauli-operator variances using Neyman-type rule \cite{arrasmith2020operator,kubler2020adaptive} or balance shots across partitioned sub-circuits using hardware error rates and the budget on total shots \cite{kulkarni2026noise}. Specifically, these papers optimize total estimation variance across circuits under uniform or mildly varying cost. These works do not consider the scenario where a feedback chain causes sequential error propagation, as is the case in IPEA. 

IPEA faces an issue of sequential error propagation  along with potentially rapidly growing iteration costs $w_k$ and noise rates $q_k$. Formulating this problem as maximization of the probability of accurate decoding of all $L$-bits  subject to a weighted resource budget $\sum_{k=1}^L w_k N_k \le W$  offers two distinct advantages. Practically, it yields an offline, pre-compiled shot schedule with zero classical runtime feedback. Mathematically, through tight upper and lower surrogate analysis via KKT conditions, it leads to a close form optimal (up to a $1-o(1)$ factor) allocation that brings out interesting resource regimes of practical interest.

\subsection{Main Contributions}

The perspective of this paper differs from the usual question of determining the total query 
complexity required to attain a prescribed phase precision and confidence. Rather, we ask 
how a fixed finite resource budget should be allocated within the procedure of the iterative 
phase estimation algorithm (IPEA) on noisy hardware of NISQ and early FT era. In particular, we consider the following practical issue up to a broad generality:  how to allocate the number of shots to different iterations of IPEA to maximize the overall accuracy (not just iteration-wise) when 
implementation cost $w_k$ of $U^{2^{k-1}}$ and the noise ($q_k$) in that circuit vary with $k$. 

As direct optimization of the exact success probability is computationally intractable, we bound it using concentration and anti-concentration inequalities to form lower- and upper-surrogate problems. In Theorem~\ref{thm:asymptotic_optimality}, we show that the accuracy achieved at the lower surrogate optimum $\hat{\mathbf{N}}$ is within a factor of $1-o(1)$, when the budget satisfies $W = \sum_{k=1}^L \frac{w_k}{c_k}(\ln L + \omega(1))$ for $c_k=-\ln\left(2\sqrt{\left(1-\frac{q_k}{2} \right) \frac{q_k}{2}}\right)$.  We then derive a closed-form for $\hat{\mathbf{N}}$  by solving the lower surrogate problem for a general iteration-dependent cost and error.

A direct practical take-away of this closed-form solution is that a total resource budget of $W = (1+r)\ln(L)\sum_{k=1}^L \frac{w_k}{c_k}$ achieves an accuracy within a $1 - L^{-r}$ factor of the theoretical optimum. Consider a realistic problem size, $L \sim 10^1$ for quantum simulations to $L\sim 10^4$ for RSA factoring, which is run on early fault-tolerant architectures  with $q_k \le 10^{-2}$. For this problem, if the total budget can allow $5(1+r)$ repetitions of each iteration, our simple closed form solution guarantees an accuracy within $1 - 10^{-r}$ of the optimum.

 Further, the closed-form solution leads to interesting practical implications for resource allocation. We observe that there are two resource regimes: resources are abundant, i.e., $W = \omega\left(\ln(L) \sum_k \frac{w_k}{c_k}\right)$ and resources are sufficient but not abundant, i.e., $W=(\ln L + \omega(1)) \sum_k \frac{w_k}{c_k}$ and $W=O\left(\ln L \sum_k \frac{w_k}{c_k}\right)$. This categorization of regimes for total budget $W$, helps in designing IPEA runs in practice.

\section{Preliminaries}
\subsection{Kitaev's Iterative Phase Estimation Algorithm with Statistical Error Correction}

    Let $U$ be a unitary operator with eigenstate $|\psi\rangle$ satisfying \( U|\psi\rangle=e^{2\pi i\phi}|\psi\rangle \), where $\phi=0.\phi_1 \phi_2 \ldots \phi_L\in[0,1)$ and $\phi_k \in \{0,1\}, 1 \le k \le L$. 

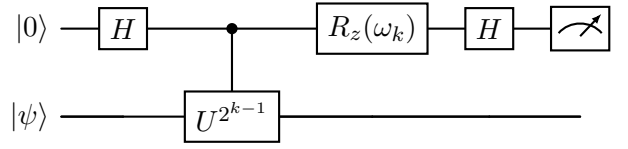
\begin{figure}[ht]
\centering
\begin{quantikz}
\lstick{$|0\rangle$}
    & \gate{H}
    & \ctrl{1}
    & \gate{R_z(\omega_k)}
    & \gate{H}
    & \meter{} \\
\lstick{$|\psi\rangle$}
    & \qw
    & \gate{U^{2^{k-1}}}
    & \qw
    & \qw
    & \qw
\end{quantikz}
\caption{$k^{\text{th}}$ iteration of Kitaev's iterative phase estimation algorithm, circuit adapted from ~\cite{dobvicek2007}.}
\label{fig:ipea}
\end{figure}

Kitaev's IPEA \cite{kitaev1995quantum} estimates $\phi$ starting with $\phi_L$ followed by $\phi_{L-1}, \phi_{L-2} \ldots \phi_1$, in that order. The circuit for estimating $k$-th phase bit, $\phi_k$, is given in Figure~\ref{fig:ipea}. It has two main quantum operations. First, application of a  $U^{2^{k-1}}$ on $|\psi\rangle$ controlled by the output $\frac{1}{\sqrt{2}}\left(|0\rangle + |1\rangle\right)$ of the first Hadamard gate. Second,  $R_z(\omega_k)$ rotation of the control qubit output of $U^{2^{k-1}}$, where the feedback phase $\omega_k = -2\pi(0.0\phi_{k+1}\phi_{k+2}\ldots \phi_L)$. Finally, the output of $R_z(\omega_k)$ is rotated by another Hadamard gate followed by a measurement in the computational basis.

In the absence of error or noise, if $\phi_L, \phi_{L-1}, \ldots, \phi_{k+1}$ have been estimated correctly, one shot of this circuit ($k$-th iteration) accurately outputs the fractional part of $2^{k-1}\phi$, and thus $\phi_k$. However, due to noise, the measured bit may or may not be exactly the fractional part of $2^{k-1}\phi$, and thus may result in wrong $\phi_k$. 
To improve reliability of this iteration (shot) is repeated $N_k$ times, and the majority is taken to decide $\phi_k$. 

The majority approach is for correcting two sources of error. (i) Given $\phi_L, \phi_{L-1}, \ldots, \phi_{k+1}$ are known correctly, there can be error in estimate of $\phi_k$ due to noise in the circuit for $k$-th iteration. (ii) Erroneous estimates of $\phi_L, \phi_{L-1}, \ldots, \phi_{k+1}$ leads to erroneous $\omega_k$, which may result in erroneous estimate of $\phi_k$ even if the $k$-th iteration is error free.

Since the error in higher-indexed iterations has an impact on lower-indexed iterations, the number of shots per iteration would ideally be different. The goal should be to choose them for maximizing overall reliability. For that,  accumulation of errors over iterations ($L$ to $1$) and its impact on overall reliability should be analyzed. 

\subsection{Modeling Error in an Iteration of IPEA}
For analyzing accumulation of errors, we need a model of error. For this work we consider NISQ and early FT hardware since they are the most relevant hardware in near future. 

In the absence of full FT, error mitigation or noise adapted error correction would lead to accumulation of errors. A tractable yet practical way to model that error would be the following. The output state before the measurement in a shot of the iteration is the correct state with some probability. Otherwise it is an unknown state with the remaining probability. When the unknown state is completely unpredictable, the noise for iteration $k$ is equivalent to $q_k$-depolarizing noise on the final quantum state (before measurement). Hence, in the rest of the paper, we shall model noise as $q_k$-depolarizing on the final state of the iteration.

\subsection{Error Accumulation over Iterations}
Let the ideal value of the $j$th bit be $\phi_j\in\{0,1\}$ and its estimate be $\hat{\phi}_j$. We define the signed estimation error $\delta_j=\hat{\phi}_j-\phi_j$, where \(\delta_j\in\{-1,0,1\} \). Since each previously estimated bit contributes to the feedback phase applied in subsequent iterations, an incorrect decision introduces a residual phase error. At the $k^\text{th}$ iteration, the accumulated phase error is obtained recursively from LSB to MSB as,

\begin{equation}
    \Delta_k = \pi \sum_{j=k+1}^{L} \frac{\delta_j}{2^{\,j-k}}.
    \label{eq:phase_error}
\end{equation}

Using this notion of the accumulated phase error \(e^{i\Delta_k}\) in the final state before measurement at the $k^{\text{th}}$ iteration, we obtain the conditional probability of correctly estimating the $k^\text{th}$ bit from the overlap of the ideal state \(|\psi_{\mathrm{ideal}}^{(k)}\rangle
= \frac{|0\rangle+(-1)^{\phi_k}|1\rangle}{\sqrt{2}}\) and the actual state \( |\psi_{\mathrm{actual}}^{(k)}\rangle = \frac{|0\rangle + (-1)^{\phi_k}e^{i\Delta_k}|1\rangle}
{\sqrt{2}}\) as,
\begin{equation}
    \PR_{\mathrm{succ}}^{(k)} = \left|\left\langle\psi_{\mathrm{ideal}}^{(k)} \middle|\psi_{\mathrm{actual}}^{(k)} \right\rangle \right|^2 = \cos^2\!\left(\frac{\Delta_k}{2}\right)
    \label{eq:success_probability}
\end{equation}

\section{Optimization Problem Formulation}

We estimate a phase $\phi=0.\phi_1\phi_2\ldots\phi_L$ through iterative phase estimation, decoding bits from LSB ($\phi_L$) to MSB ($\phi_1$). Let $E_k$ denote the event that bit $k$ is decoded incorrectly by majority vote over $N_k$ repeated single-shot measurements. As discussed before, our primary objective in this paper is to maximize the probability that all $L$ bits are correctly estimated. Mathematically,

    \begin{equation}
    \max_{\{N_k\}} \PR\big(E_1^c \cap E_2^c \cap \cdots \cap E_L^c\big)
     \text{ s.t. } \sum_{k=1}^L N_k w_k \le W.
    \label{eq:case1-obj}
    \end{equation}

The number of measurement shots $N_k$ is inherently a discrete, non-negative integer value. However, while optimizing over a discrete space for large $L$, we treat $N_k$ as continuous, non-negative real variable. Once the exact optimal continuous allocation $N_k^\ast$ is derived, the practical repetitions count to be executed on hardware is obtained by rounding to  $N_k' = \lfloor N_k^\ast \rfloor$, to strictly satisfy the budget constraint $\sum N_k' w_k \le W$. When $W\ge L$, which is the practical case to consider (as we require $W \ge \sum_k w_k \ge L$ for no iterations to be skipped), the rounding errors are negligible. 

The optimization problem aims to maximize the exact success probability of estimating the entire bit-string. Let $E_k^c$ denote the event that bit $k$ is decoded correctly by majority vote over $N_k$ repeated single-shot measurements, so the objective function becomes \(\PR\big(E_1^c \cap E_2^c \cap \cdots \cap E_L^c\big)\). Using the product rule, we have,
\begin{align}
\PR(E_1^c\cap\cdots\cap E_L^c) = \prod_{k=1}^{L} \PR \big(E_k^c \mid E_{k+1}^c\cap\cdots\cap E_L^c\big).
\label{eq:chain-rule}
\end{align}

For evaluating $\prod_{k=1}^{L} \PR \big(E_k^c \mid E_{k+1}^c\cap\cdots\cap E_L^c\big)$, we condition on all previously decided bits being correct. Consequently, there is no residual phase error, i.e., we have  accumulation of error at $k^\text{th}$ iteration given $E_{k+1}^c\cap\cdots\cap E_L^c$ is $0$, i.e., $\Delta_k = 0$.

Applying the depolarizing noise channel model, the single-shot success probability conditioned on correct prior decoding \( \PR(\text{success in one shot of iteration } k \mid E_{k+1}^c\cap\cdots\cap E_L^c) = \frac{q_k}{2} + (1-q_k)\cos^2(0) = 1 - \frac{q_k}{2} \equiv p_k. \) 
Though there can be accumulation of errors in IPEA, the expression for success in one shot of $k$ given all previous bits are correct decouples. However, obtaining an exact expression for $\PR \big(E_k^c \mid E_{k+1}^c\cap\cdots\cap E_k^c\big)$ from this and optimizing that expression is intractable. Hence, we obtain upper and lower bounds on this probability using concentration inequalities  \cite{boucheron2013concentration} and anti-concentration inequalities based on method of types \cite[Ch.~11]{cover2006elements}.

\section{Surrogate bounds}

To bound the conditional probability for each bit, let $X_1, X_2, \ldots, X_{N_k} \in \{0,1\}$ be independent and identically distributed (i.i.d.) indicator variables for correct single-shot outcomes, where $X_i \sim \text{Ber}(p_k)$. The expected number of successful shots is $\EX\big[\sum_{i=1}^{N_k}X_i\big] = N_k p_k$.

The majority vote decoding fails if the number of correct outcomes is less than or equal to half the total shots $\sum_{i=1}^{N_k}X_i \leq \frac{N_k}{2}$. 
By concentration inequalities  \cite{boucheron2013concentration} and anti-concentration inequality based on method of types bounds \cite[Ch.~11]{cover2006elements} for Bernoulli random variables, probability of failure is upper and lower bounded by

$$\exp(-c_k \cdot N_k) \text{  and  } \frac{1}{(N_k+1)^2} \exp(-c_k \cdot N_k),$$ respectively, for even $N_k$ where,

$$c_k=-\ln\left(2\sqrt{p_k(1-p_k)}\right)>0,$$
if $q_k<1$. Also, $c_k \to \infty$ whenever $q_k \to 0$.

A tighter lower bound could be obtained for all $N_k$ using the Bahadur-Rao bound \cite{bahadur1960deviations,dembo2010large}. However, as we show in Theorem~\ref{thm:asymptotic_optimality}, that the above bound is  sufficient to lead to an optimal close form allocation. From the proof of Theorem~\ref{thm:asymptotic_optimality}, it would be clear that the Bahadur-Rao bound does not change the significant terms of the allocation.

Using the above upper and lower bound on the probability of failure we get 
\begin{align}
&1-\frac{1}{(N_k+1)^2} \exp(-c_k \cdot N_k) \nonumber \\  & \ge \PR\big(E_k^c \mid E_{k+1}^c\cap\cdots\cap E_L^c\big) \ge 1-\exp(-c_k \cdot N_k).  
\label{eq:hoeffding}
\end{align}

Thus, the success probability is, respectively, upper and lower bound by

\begin{align}
    & \prod_{k=1}^L \left(1-\frac{1}{(N_k+1)^2} \exp(-c_k \cdot N_k)\right)  \nonumber \\
    &~~~~~~~~~\text{and }  \prod_{k=1}^L \left(1- \exp(-c_k \cdot N_k)\right).
\end{align}
 
Therefore, the following optimization problem gives a surrogate (lower bound) for the success probability maximization problem:
\begin{equation}
\max_{\{N_k \ge 0\}} \; \sum_{k=1}^{L} \ln\!\big(1-e^{-c_k N_k}\big)
\quad \text{s.t.} \quad \sum_{k=1}^L N_k w_k \le W.
\label{eq:ln}
\end{equation}

Similarly, one can also consider the upper bound as a surrogate. We discuss this case later and show that we obtain similar results from both when the resource is not severely constrained and can scale with $L$.

We denote the surrogate optimal allocations of the lower surrogate function \(f_{\text{LS}} = \sum_{k=1}^L \ln(1-e^{-c_k N_k})\)  as  $\hat{\mathbf{N}}$, with Lagrange multiplier $\lambda_{\text{LS}}$ and of the upper surrogate function \(f_{\text{US}} = \sum_{k=1}^L \ln \left( 1-\frac{1}{(N_k +1)^2}e^{-c_k N_k}\right)\) as $\tilde{\mathbf{N}}$, with Lagrange multiplier $\lambda_{\text{US}}$.

To make the derivative simpler, we convert the product of exponentials to a summation using the natural logarithm. Because $\ln$ is strictly increasing and strictly concave, the transformed objective function $\sum_{k=1}^{L} \ln(1-e^{-c_kN_k})$ is jointly concave on a convex constraint set. The inequality constraint $\sum_{k=1}^L N_k w_k - W \le 0$ is affine. Choosing any strictly positive uniform allocation $\hat{\mathbf{N}}_k > 0$ that satisfies the budget constraint with strict inequality demonstrates that Slater's condition holds \cite{boyd2004convex}. Consequently, strong duality is satisfied, proving that KKT stationary conditions are both necessary and sufficient for identifying the unique global maximizer.

In the practical case of $W\gg L$, the optimal solution of upper and lower surrogate would ensure ${\mathbf{N}}_k>0$ since the objective function $\to -\infty$ when $N_k\to 0$. Hence, we can drop the condition $N_k\ge 0$ from the Lagrangian without changing the solution of the optimization problems when we are in the practically useful range of $W \gg L$.

\subsection{Lagrangian and KKT Conditions (General Formula)}
\subsubsection{Lower-bound surrogate problem}
Let us consider the Lagrangian and KKT conditions for the lower bound surrogate.
For the inequality constraint $\sum_k N_k w_k - W \le 0$, we form the Lagrangian for \( \lambda_{\text{LS}} \ge 0\),
\begin{equation}
\mathcal{L} = \sum_{k=1}^L \ln\!\big(1-e^{-c_k N_k}\big)
- \lambda_{\text{LS}}\Big(\sum_{k=1}^L N_k w_k - W\Big). \nonumber
\end{equation}

Applying the KKT conditions gives a generalization for the optimal allocation $\hat{\mathbf{N}}$,
\begin{equation}
    \hat{\mathbf{N}}(\lambda_{\text{LS}}) = \frac{1}{c_k}\ln\!\left(1+\frac{c_k}{\lambda_{\text{LS}} w_k}\right).
    \label{eq:dep_sol}
\end{equation}   

To solve for $\lambda_{\text{LS}}$ we use the fact that, the optimal solution for \eqref{eq:dep_sol} satisfies $\sum_k \hat{\mathbf{N}} w_k = W$.

\begin{equation}
    \sum_{k=1}^L\frac{w_k}{c_k}\ln\!\left(1+\frac{c_k}{\lambda_{\text{LS}} w_k}\right)=W.
\end{equation}

\subsubsection{Upper-bound surrogate problem}
In this case, the steps are similar, but due to the extra multiplicative term $\frac{1}{(\tilde{N}_k+1)^2}$ inside the log, the calculations of the derivatives are a bit tedious. We finally obtain

$$\frac{c_k(\tilde{N}_k+1) + 2}{(\tilde{N}_k+1)\Big((\tilde{N}_k+1)^2 e^{c_k \tilde{N}_k} - 1\Big)} = \lambda_{\text{US}} w_k.$$

We can obtain $\lambda_{\text{US}}$ by solving the equation $\sum_k \tilde{N}_k w_k = W$. Thus, the solution is structurally similar to the lower-bound surrogate case.

\subsubsection{Closeness of upper and lower surrogate for $q_k \ll 1$}
For any quantum circuit to work, in practice $q_k\ll 1$ and hence, $1-p_k \ll 1$. Hence,  $c_k\gg 1$ for almost all practically useful circuits. When $c_k \gg 1$,  from the upper-bound surrogate problem, we get

$$(\tilde{N}_k+1)^2 e^{c_k \tilde{N}_k} - 1 = \frac{c_k}{\lambda_{\text{US}} w_k},$$

which implies
$$\tilde{N}_k + \frac{2}{c_k} \ln(\tilde{N}_k+1)=\frac{1}{c_k}\ln(1+\frac{c_k}{\lambda_{\text{US}} w_k}).$$

Since $c_k\gg 1$, we can ignore $\frac{2}{c_k} \ln(\tilde{N}_k+1)$ compared to $\tilde{N}_k$. Hence, the solution of the upper surrogate problem becomes the same as that of lower surrogate problem in the practically useful case of $q_k\ll 1$ (i.e., $c_k \gg 1$). As a result, the solution of the lower surrogate problem is close to the true optimum.

We state a formal result along this line in Theorem~\ref{thm:asymptotic_optimality}. In particular, the theorem shows that the accuracy achieved at the optimizer of the lower surrogate is arbitrarily close to the optimum accuracy. This formal result is applicable for any $q_k \in [0,1)$.

\subsection{Closeness between Optima of True Objective and Lower Surrogate}

While $\PR(\text{all } L \text{ bits correct})$ represents the exact success probability of IPEA, direct optimization of this is computationally intractable. To validate using lower-bound surrogate in place of true objective, we establish an explicit bound on the gap between the true optimal allocation and our closed-form lower surrogate allocation.

\begin{theorem}[Asymptotic Optimality Guarantees for Surrogate IPEA Allocation]
\label{thm:asymptotic_optimality}
Let $g(\{N_k\})=\mathbb{P}(\text{succ})$ denote the true, exact success probability of the IPEA algorithm for all $L$ bits correct under allocation vector $\mathbf{N} = (N_1, \dots, N_L)$. It is bounded point-wise by $\ \prod_{k=1}^L  (1-e^{-c_k N_k}) \le g(\mathbf{N}) \le \prod_{k=1}^L  \left(1-\frac{e^{-c_k N_k}}{(N_k+1)^2}\right)$. Let $\mathbf{N}_k^*$ be the intractable true optimal allocation of the optimization problem maximizing $P(\text{succ})$ subject to $\sum_{k=1}^L N_k w_k \leq W$. Let $\hat{\mathbf{N}_k}$ and $\tilde{\mathbf{N}_k}$ be the closed-form optimal allocation derived from optimizing the lower-bound surrogate function $f_{\text{LS}} = \sum_{k=1}^L \ln(1-e^{-c_k N_k})$ and the upper bound surrogate function $f_{\text{US}}=\sum_{k=1}^L \ln\left(1-\frac{e^{-c_k N_k}}{(N_k+1)^2}\right)$ respectively. Then, the true success probability achieved by lower surrogate allocation $\hat{\mathbf{N}}$ satisfies the following.

\begin{equation}
 \ln g(\mathbf{N}^*) - \ln g(\hat{\mathbf{N}}) = O\left(L \exp\left( -\frac{W}{\sum_{k=1}^L \frac{w_k}{c_k}} \right)\right). \nonumber
\end{equation}

If the total budget satisfies the scaling condition $W = \sum_{k=1}^L \frac{w_k}{c_k} \big( \ln(L) + \eta\big)$ for $\eta>0$, then:
\begin{equation}
    g(\mathbf{N}^*) \;\le\; (1 + e^{-\eta}) g(\hat{\mathbf{N}}) \mbox{, and } \nonumber
\end{equation}
and thus $g(\mathbf{N}^*) \;\le\; (1 + o(1)) g(\hat{\mathbf{N}})$ when $\eta=\omega(1)$.

\end{theorem}

The full proof of Theorem~\ref{thm:asymptotic_optimality} is provided in Appendix~\ref{proof_theorem}. This guarantees that optimizing the tractable lower surrogate $f_{\text{LS}}$, which is analytically more tractable, yields a true success probability that approaches the optimal success probability $g(\mathbf{N_k}^*)$ whenever $W$ is sufficiently large.

We observe that the analytical guarantee in Theorem~\ref{thm:asymptotic_optimality} is useful only when $W\ge\ln(L) \sum_{k=1}^L \frac{w_k}{c_k}$. More specifically, if $W=(\ln(L)+\eta(L)) \sum_{k=1}^L \frac{w_k}{c_k}$, then the allocation of shots based on the optimization of the lower surrogate achieves an accuracy within $1-e^{-\eta(L)}$ factor of the optimum. Thus, if $W=(1+r)\ln(L) \sum_{k=1}^L \frac{w_k}{c_k}$, then the achieved accuracy is within a factor of $1-\frac{1}{L^r}$ of the optimum accuracy.    
The condition $W=(1+r)\ln(L) \sum_{k=1}^L \frac{w_k}{c_k}$ is satisfied when $W \ge (1+r) \frac{\ln L}{\min_k c_k} \sum_k w_k$. For many applications ranging from quantum simulations to factoring (RSA) $L \sim 10^1$ -- $10^4$ and hence, $\ln L \le 10$. Since in early fault-tolerant systems $q_k \le 10^{-2}$  (and hence, $c_k\gtrsim 2$), the condition $W = (1+r) \frac{\ln L}{\min_k c_k} \sum_k w_k$ becomes equivalent to: $W$ is no less than $5(1+r)\sum_k w_k$. That means if there is enough total resource to run each iteration $5(1+r)$ times, the allocation of shots obtained by solving the lower surrogate problem achieves an accuracy which is within a factor of $1-{10^{-r}}$ of the optimum accuracy. 

 \section{Accuracy maximizing allocation of shots}
In this section, building on the guarantee in Theorem~\ref{thm:asymptotic_optimality}, we obtain closed form expression for $\left(1-o(1)\right)$-optimal shots allocation by solving for $\{\hat{N}_k\}$ when $W=\sum_k \frac{w_k}{c_k} (\ln L+\omega(1))$.

We start with some special cases of $w_k$ and $c_k$ motivated by IPEA for order finding, which has multiple applications. After that we discuss the case for general $w_k$ and $c_k$, which is important for quantum simulations.

\subsection{Special $w_k$ and $c_k$: applications in order-finding}
\label{sec:special}
In the order-finding problem \cite{nielsenchuang}, thanks to the modular arithmetic circuit, the complexity of circuit remains uniform across iterations. Since the circuits for each iteration has similar structure,  $w_k=w$ and $c_k=c$ for all $k$ is the first natural scenario to consider. The solution is simple in this case due to symmetry:

\begin{equation}
    \hat{\mathbf{N}}_k(\lambda_{\text{LS}}) = \frac{1}{c}\ln\!\left(1+\frac{c}{\lambda_{\text{LS}} w}\right) \text{ and }
\end{equation}

\begin{equation}
    \sum_{k=1}^{L} w\hat{\mathbf{N}} = W \Rightarrow \hat{\mathbf{N}}=\frac{W}{L~w}.
\end{equation}

Next, we consider another scenario motivated by early fault-tolerant IPEA for order-finding: $w_k$ grows linearly with $c_k$. This considers the case where the circuit complexity (cost) for the $k$-th iteration grows linearly with the error mitigation cost, captured by $c_k$, which is a function of $q_k$. Since $c_k$ is proportional to $\ln\frac{1}{q_k}$ for $q_k\ll 1$, this scenario is same as $w_k$ scaling linearly with $\ln\frac{1}{q_k}$. This case represents the scenario where the circuit complexity of different iterations are the same; however, different iterations can have different error mitigation techniques (representing different $q_k$) and the resource cost increases with the error mitigation requirement. The particular dependence of the cost for error mitigation on $\ln\frac{1}{q_k}$ captures the amount of redundancy needed for achieving $1-q_k$ accuracy. This is because to achieve a logical accuracy of $1-q_k$ with $q_k \ll 1$, the space-overhead ($1/$code rate) needs to scale as $\ln\frac{1}{q_k}$, when physical errors are i.i.d.

Given $\beta = c_k/w_k$ for all $k$ the general allocation reduces to
\begin{equation}
    \hat{\mathbf{N}}_k(\lambda_{\text{LS}})  = \frac{1}{c_k}\ln{\left( 1+ \frac{\beta}{\lambda_{\text{LS}}}\right)}
\end{equation}

This simplifies further when plugged back into the budget constraint equation,

\begin{align}
    \sum_{k=1}^{L} w_k\hat{\mathbf{N}}_k = \sum_{k=1}^{L} \frac{w_k}{c_k}\ln{\left( 1+ \frac{c_k}{\lambda_{\text{LS}} w_k}\right)}&= W \nonumber \\
      \sum_{k=1}^{L} \frac{1}{\beta}\ln{\left( 1+ \frac{\beta}{\lambda_{\text{LS}}}\right)} = \frac{L}{\beta}\ln{\left( 1+ \frac{\beta}{\lambda_{\text{LS}}}\right)} &= W \nonumber \\
   \hat{\mathbf{N}}_k w_k = \frac{1}{\beta}\ln{\left( 1+ \frac{\beta}{\lambda_{\text{LS}}}\right)}  &= \frac{W}{L}  \nonumber
\end{align} 

This gives us the optimal allocation $\hat{\mathbf{N}}_k$ and $\lambda_{\text{LS}}$ as,
\begin{equation}
    \hat{\mathbf{N}}_k =\frac{W}{L~w_k} \text{ and, } \lambda_{\text{LS}} = \frac{\beta}{e^\frac{W\beta}{L} - 1}
\end{equation}

\subsection{General $w_k$ and $c_k$: applications in quantum simulations}

In practical implementations of Kitaev's IPEA for quantum simulations, since estimating lower significant bits requires passing it through the $U^{2^{k-1}}$ gate it increases the circuit depth for lower significant bits, thereby increasing the cost $w_k$. The way $w_k$ grows with $k$ depends on the structure of the particular unitary operation.
 Here, we consider the case where $w_k$ and $c_k$ scale arbitrarily and obtain an expression for $\hat{\mathbf{N}}$.

When \(W >  \sum_k \frac{w_k}{c_k} \ln L\), by Theorem~\ref{thm:asymptotic_optimality}, the expression $\hat{\mathbf{N}}(\lambda_{\text{LS}}) = \frac{1}{c_k}\ln\!\left(1+\frac{c_k}{\lambda_{\text{LS}} w_k}\right)$ is valid. However, it is intractable to solve for $\hat{N}_k$. We make an assumption that the Lagrange multiplier $\lambda_{\text{LS}}$ is sufficiently small so that \(\frac{c_k}{\lambda_{\text{LS}} w_k} \gg 1\), derive $\lambda_{LS}$ and show that the assumption is self-consistent.

When \(\frac{c_k}{\lambda_{\text{LS}} w_k} \gg 1\), using the fact that $\ln(1+x) \approx \ln(x)$ for $x\gg 1$, we   obtain,

\begin{equation}
    \hat{\mathbf{N}}(\lambda_{\text{LS}}) = \frac{1}{c_k}\ln\!\left(1+\frac{c_k}{\lambda_{\text{LS}} w_k}\right) \approx \frac{1}{c_k}\ln\!\left(\frac{c_k}{\lambda_{\text{LS}} w_k}\right) 
    \label{eq:N_k approx}
\end{equation}

Substituting \eqref{eq:N_k approx} into the budget constraint equation \( \sum_{k=1}^L w_k \hat{\mathbf{N}}_k = W\), we obtain,

\begin{equation}
    W  = \sum_{k=1}^L \frac{w_k}{c_k} \left[ \ln\left(\frac{c_k}{w_k}\right) + \ln\left(\frac{1}{\lambda_{\text{LS}}}\right) \right].
\end{equation}

Isolating the term $\lambda_{\text{LS}}$, 
\begin{equation}
    \frac{1}{\lambda_{\text{LS}}}  = \exp \left( \frac{W + \sum_{k=1}^L \frac{w_k}{c_k} \ln\left(\frac{w_k}{c_k}\right)}{\sum_{k=1}^L \frac{w_k}{c_k}} \right).
    \label{eq:lambda_exp}
\end{equation}

Since we have established that $\lambda_{\text{LS}}$ scales exponentially, the assumption that $\frac{c_k}{\lambda_{\text{LS}} w_k} \gg 1$ holds strictly provided that $\frac{c_k}{w_k}$ does not grow faster than exponentially in $\frac{W}{\sum_{k=1}^L \frac{w_k}{c_k}}$. This is further proved in Lemma \ref{lem:ckwk_condition} in Appendix \ref{lemmas}. 

Substituting \eqref{eq:lambda_exp} back into \eqref{eq:N_k approx}, we obtain a closed form:
\begin{equation}
    \hat{N}_k = \frac{1}{c_k} \left[ \ln\left(\frac{c_k}{w_k}\right) + \frac{W + \sum_{k=1}^L \frac{w_k}{c_k} \ln\left(\frac{w_k}{c_k}\right)}{\sum_{k=1}^L \frac{w_k}{c_k}} \right]. \nonumber
    \label{eq:general_Nk}
\end{equation}

As discussed above, though $\hat{N}_k$ maximizes a lower bound on the accuracy, by Theorem~\ref{thm:asymptotic_optimality}, when phase estimation  is run with  $N_k=\hat{N}_k$ the achieved accuracy is within $1-o(1)$ factor of the optimum accuracy.

\subsubsection{Practical implications: resource regimes}

For many applications, either $c_k$ and $w_k$ are known or can be estimated from the circuit and hardware parameters. Theorem~\ref{thm:asymptotic_optimality} implies that repeating iteration $k$ according to the above expression for $\hat{N}_k$ would optimize the performance of IPEA.

The expression for $\hat{N}_k$ can be seen as $\frac{R(W,\{w_k,c_k\})}{c_k}  - \frac{\ln w_k - \ln c_k}{c_k}$, where $R(W,\{w_k,c_k\})=\left(W + \sum_{k=1}^L \frac{w_k}{c_k} \ln\left(\frac{w_k}{c_k}\right)\right)/{\sum_{k=1}^L \frac{w_k}{c_k}}$ depends on the overall resource cost and constraints, and is the same for all iteration $k$. 

When $R\gg \max_k \vert \ln(w_k/c_k)\vert$,  the optimal shots allocation becomes: iteration $k$ is allocated $\frac{R(W,\{w_k,c_k\})}{c_k}$ shots. In that case, the optimal allocation depends only on individual $c_k$ and the system-level quantity $R$. Thus, even when $R$ is not known exactly,  shots can be allocated optimally by ensuring they are proportional to $\frac{1}{c_k}$. 

Using  Jensen's inequality to the convex function $x\ln x$ for $x>0$, it turns out that a sufficient condition for $R\gg \max_k \vert \ln(w_k/c_k) \vert$ is:

$$\frac{W}{\sum_{k=1}^L \frac{w_k}{c_k}} \gg  \left[\ln L - \ln\left(\sum_{k=1}^L \frac{w_k}{c_k}\right) + \max_{1 \le k \le L} \vert{}\ln \frac{w_k}{c_k}\vert{}\right]$$

The scaling of $\frac{w_k}{c_k}$ with $k$ depends on the particular phase estimation problem. If the unitary has no special structure to exploit, as often the case in certain quantum simulations, $w_k$ would increase exponentially with $k$. 
In general, if no error correction is applied, $c_k$ would decrease exponentially with $k$ since the depth of the circuit increases exponentially. However, when error correction or error mitigation techniques are applied, the scaling of $c_k$ would be slower. Thus, $\frac{w_k}{c_k}$ would increase exponentially, dominated by the scaling of $w_k$.
On the other hand, when the unitary used in the quantum simulation has an exploitable structure, the scaling of $w_k$ and hence, that of $\frac{w_k}{c_k}$ can be polynomial.

When $\frac{w_k}{c_k}$ increases polynomially or faster with $k$, which, as discussed above, is the case for many quantum simulation scenarios, $\ln\left(\sum_{k=1}^L \frac{w_k}{c_k}\right) \sim \max_{1 \le k \le L} \vert{}\ln \frac{w_k}{c_k}\vert$. Thus, in those settings, if $W \gg \ln(L) \sum_k \frac{w_k}{c_k}$, the optimal shot allocation is simple: iteration $k$ is allocated shots proportional to $\frac{1}{c_k}$.

In Theorem~\ref{thm:asymptotic_optimality}, the condition for $\hat{\mathbf{N}}$ to be $1-o(1)$ optimal is $W=(\ln(L)+\omega(1))\sum_k \frac{w_k}{c_k}$. On the other hand, for the above simple allocation policy (proportional to $\frac{1}{c_k}$) to be applicable, we should have order-wise more resource: $W= \omega\left( \ln(L)\cdot \sum_k \frac{w_k}{c_k}\right)$.

When $(\ln(L)+\omega(1))\sum_k \frac{w_k}{c_k}\le W \le O\left(\ln(L)\cdot \sum_k \frac{w_k}{c_k}\right)$, the simple allocation rule (proportional to $\frac{1}{c_k}$) is not optimal and we have to use the general formula for $\hat{\mathbf{N}}$ derived above. This general formula can be seen as an additive correction to the simple allocation by  $-\frac{1}{c_k}\ln\frac{w_k}{c_k}$.

The probability of error in deciding $k^\text{th}$ bit after majority decision based on $\hat{N}_k$ shots is approximately $\exp(-c_k \hat{N}_k)$. Thus, when $W= \omega\left(\ln(L)\cdot \sum_k \frac{w_k}{c_k}\right)$, the optimal allocation rule can also be seen as: allocate shots such that each bit has the same probability of error given by $e^{-R(W,\{w_k,c_k\})}$. This means that when there is abundant resource, i.e., $W= \omega\left(\ln(L)\cdot \sum_k \frac{w_k}{c_k}\right)$, the optimal resource allocation ensures that each bit achieves the same accuracy.

On the other hand, when $(\ln(L)+\omega(1))\sum_k \frac{w_k}{c_k}\le W \le O\left(\ln(L)\cdot \sum_k \frac{w_k}{c_k}\right)$, under the optimal allocation, for any $k$, the probability of error of the  $k$-bit conditioned on the less-significant bits ($k+1$ through $L$) being correct, when multiplied by $\frac{c_k}{w_k}$ is identical for all $k$, i.e., $\frac{c_k}{w_k} P_{\text{err}}(k) \approx e^{-R(W,\{w_k,c_k\})}$. In comparison with the MSB ($k=1$), it implies that, $P_{\text{err}}(k)=\frac{1}{\frac{c_k w_1}{c_1 w_k}}P_{\text{err}}(1).$ This means that when the resource is sufficient but not abundant, the optimal allocation ensures that the probability of error of the lower significant bit $k$ is lower by a factor of $\frac{c_k w_1}{c_1 w_k}$ compared to the most significant bit. 

Thus, when the resource is not abundant the optimum allocation has to account for the fact that in IPEA error propagates from lower to higher significant bits. However, when the resource is abundant, treating each bit separately is optimal.

\section{Conclusion}

Motivated by practical issues with implementation of IPEA, a popular iterative algorithm  well suited for phase estimation on NISQ and early FT hardware,   we consider the problem of allocation of number of shots across iterations. Different iterations of IPEA consume different amounts of resource in terms of circuit complexity, oracle call, etc., depending on the particular phase estimation problem, e.g., phase estimation for order finding versus that for different quantum simulations. Similarly, depending on the circuit realization of different iterations and the error mitigation schemes, the accuracy of iterations also vary. Since in NISQ and early FT circuits, iterations are not reliable, repetitions of an iteration (shots) followed by majority decision is a standard practice for  improving reliability.

We find the optimal allocation of number of shots (repetitions) per iteration that maximizes the probability of success of the overall IPEA algorithm, given a budget for the total resource. We derive closed-form expressions for optimal number of shots for different scalings of cost and reliability with iterations. We do so by obtaining upper and lower surrogates for the objective by invoking tight concentration and anti-concentration bounds and establishing an asymptotic optimality guarantee showing that optimizing the lower surrogate achieves full-phase accuracy close to the true optimum once the total budget exceeds a logarithmic threshold in the number of phase bits relative to the cost-to-reliability ratio across iterations.

We observe that in the resource-abundant regime (high resource budget), the allocation only depends on the reliability of the iteration. In this regime, allocating shots inversely to each iteration's noise parameter gives every bit the same error probability. However, when resources are sufficient but not abundant, the optimal allocation adds a cost- and noise-dependent adjustment to account for error propagation from less-significant to more-significant bits.

    {\bf Acknowledgment}: AC thanks ANRF India (formerly, SERB India) for support through grant CRG/2023/005345 and the Ministry of Education, India.

% \clearpage
\onecolumn
\appendix

\section{Appendix}

  Here, we show the complete proof of Theorem~\ref{thm:asymptotic_optimality} using two supporting lemmas, Lemma~\ref{lemma:surrogate_gap} and \ref{lem:UBtildeNk}. We also state and prove a third lemma, Lemma~\ref{lem:ckwk_condition} that is useful for deriving a close form expression for the solution of the lower surrogate problem.

\subsection{Proof of Theorem~\ref{thm:asymptotic_optimality}}
\label{proof_theorem}

\begin{proof}
By definition of the surrogate optimization problems, we have,
\begin{equation}
\hat{\mathbf{N}} = \arg\max_{\sum w_k N_k \le W} f_{\text{LS}}(\mathbf{N}) \quad \text{and} \quad \tilde{\mathbf{N}} = \arg\max_{\sum w_k N_k \le W} f_{\text{US}}(\mathbf{N}). \nonumber
\end{equation}

Where, \(f_{\text{LS}} = \sum_{k=1}^L \ln(1-e^{-c_k N_k})\)  and \(f_{\text{US}} = \sum_{k=1}^L \ln \left( 1- \frac{1}{(N_k +1)^2}e^{-c_k N_k}\right)\).

The point-wise bound has established that,
\begin{equation}
    \prod_{k=1}^L  (1-e^{-c_k N_k}) \le g(\mathbf{N}) \le \prod_{k=1}^L  (1-\frac{e^{-c_k N_k}}{(N_k+1)^2}) \nonumber
\end{equation}

Taking the natural logarithm of the point-wise bounds on $g(\mathbf{N})$, we establish that for any feasible allocation satisfying the budget constraint \( \sum_{k=1}^L w_k N_k =W\) we have,
\begin{equation}
f_{\text{LS}}(\mathbf{N}) \;\le\; \ln g(\mathbf{N}) \;\le\; f_{\text{US}}(\mathbf{N}). \label{eq:pointwise_log_bounds}
\end{equation}

Since $\tilde{\mathbf{N}}$ maximizes $f_{\text{US}}(\mathbf{N})$ over the feasible set $\sum_{k=1}^L w_k N_k \le W$, it follows that \(f_{\text{US}}(\mathbf{N}^*) \;\le\; f_{\text{US}}(\tilde{\mathbf{N}})\).

Starting with the true optimal performance $\ln g(\mathbf{N}^*)$ and applying the upper point-wise bound from \eqref{eq:pointwise_log_bounds}, we have
\begin{equation}
\ln g(\mathbf{N}^*) \;\le\;  f_{\text{US}} (\mathbf{N}^*) \;\le\; f_{\text{US}}(\tilde{\mathbf{N}}). \nonumber
\end{equation}

We can rewrite $f_{\text{US}}(\tilde{\mathbf{N}})$ as $f_{\text{LS}}(\tilde{\mathbf{N}}) + \left[ f_{\text{US}}(\tilde{\mathbf{N}}) - f_{\text{LS}}(\tilde{\mathbf{N}}) \right]$, to establish
\begin{equation}
\ln g(\mathbf{N}^*) \;\le\; f_{\text{LS}}(\tilde{\mathbf{N}}) + \left( f_{\text{US}}(\tilde{\mathbf{N}}) - f_{\text{LS}}(\tilde{\mathbf{N}}) \right). \nonumber
\end{equation}

Also, since $\hat{\mathbf{N}}$ maximizes $f_{\text{LS}}(\mathbf{N})$ over the feasible set $\sum_{k=1}^L w_k N_k \le W$, we have that \(f_{\text{LS}}(\tilde{\mathbf{N}}) \;\le\; f_{\text{LS}}(\hat{\mathbf{N}})\).
This gives us,

\begin{equation}
\ln g(\mathbf{N}^*) \;\le\; f_{\text{LS}}(\hat{\mathbf{N}}) + \left( f_{\text{US}}(\tilde{\mathbf{N}}) - f_{\text{LS}}(\tilde{\mathbf{N}}) \right). \nonumber
\end{equation}

Finally, applying the lower point-wise bound from \eqref{eq:pointwise_log_bounds} at $\hat{\mathbf{N}}$, which states $f_{\text{LS}}(\hat{\mathbf{N}}) \le \ln g(\hat{\mathbf{N}})$, we have,
\begin{equation}
\ln g(\mathbf{N}^*) \;\le\; \ln g(\hat{\mathbf{N}}) + \left( f_{\text{US}}(\tilde{\mathbf{N}}) - f_{\text{LS}}(\tilde{\mathbf{N}}) \right).  \nonumber
\end{equation}

By Lemma \ref{lemma:surrogate_gap},  the surrogate gap term \(f_{\text{US}}(\tilde{\mathbf{N}}) - f_{\text{LS}}(\tilde{\mathbf{N}})\) at allocation $\tilde{\mathbf{N}}$ satisfies:
\begin{equation}
f_{\text{US}}(\tilde{\mathbf{N}}) - f_{\text{LS}}(\tilde{\mathbf{N}}) \;\le\; O \left( L \exp\left( -\frac{W}{\sum_{k=1}^L \frac{w_k}{c_k}} \right) \right). \nonumber
\end{equation}

Thus, we finally have,
\begin{equation}
    \ln g(\mathbf{N}^*) \;\le\; \ln g(\hat{\mathbf{N}}) + O \left( L \exp\left( -\frac{W}{\sum_{k=1}^L \frac{w_k}{c_k}} \right) \right). \nonumber
\end{equation}

Under the asymptotic budget scaling condition $W = \sum_{k=1}^L \frac{w_k}{c_k} \big( \ln(L) + \omega(1) \big)$, we have,
\begin{align}
    \ln g(\mathbf{N}^*) - \ln g(\hat{\mathbf{N}}) &\le O 
    \left( L \cdot \exp\left( -\left( \ln(L) + \omega(1) \right) \right) \right)\nonumber \\
    &= O\left( L \cdot \frac{1}{L} \cdot e^{-\omega(1)} \right) = O( e^{-\omega(1)}) \nonumber
\end{align}

Thus, the log-probability ratio satisfies:
\begin{equation}
\ln\left( \frac{g(\mathbf{N}^*)}{g(\hat{\mathbf{N}})} \right) \le O(e^{-\omega(1)})\;  \Rightarrow \;\frac{g(\mathbf{N}^*)}{g(\hat{\mathbf{N}})} \leq e^{O(e^{-\omega(1)})}. \nonumber
\end{equation}

Applying taylor series approximation to $e^{O(e^{-\omega(1)})}$, we have  $e^{O(e^{-\omega(1)})} = 1 + O(e^{-\omega(1)}) = 1+ o(1)$

\begin{equation}
\frac{g(\mathbf{N}^*)}{g(\hat{\mathbf{N}})} \le 1 + o(1) \implies g(\mathbf{N}^*) \le (1 + o(1)) \, g(\hat{\mathbf{N}}). \nonumber
\end{equation}

This confirms that the explicit lower-surrogate allocation policy $\hat{\mathbf{N}}$ achieves asymptotic optimality relative to the exact optimal allocation $\mathbf{N}^*$.

\end{proof}

\subsection{Lemmas}
\label{lemmas}

\begin{lemma}
\label{lemma:surrogate_gap}
Let $f_{\text{US}}(\tilde{\mathbf{N}})$ and $f_{\text{LS}}(\hat{\mathbf{N}})$ be the optimal values of the upper and lower surrogate objectives. In the high-budget regime ($W \gg \sum_{k=1}^L \frac{w_k}{c_k}$), the objective gap $ f_{\text{US}}(\tilde{\mathbf{N}}) - f_{\text{LS}}(\tilde{\mathbf{N}})$ is bounded in closed form by
\begin{equation}
f_{\text{US}}(\tilde{\mathbf{N}}) - f_{\text{LS}}(\tilde{\mathbf{N}}) \le O\left(L \exp\left( -\frac{W}{\sum_{k=1}^L \frac{w_k}{c_k}} \right) \right). \nonumber
\end{equation}
Furthermore, a budget scaling rule of $\frac{W}{\sum_{k=1}^L w_k/c_k} = \ln(L) + \omega(1)$ is sufficient to guarantee that the surrogate objective gap vanishes asymptotically, i.e.,
\begin{equation}
f_{\text{US}}(\tilde{\mathbf{N}}) - f_{\text{LS}}(\tilde{\mathbf{N}}) \le o(1). \nonumber
\end{equation}
\end{lemma}

\begin{proof}

The objective gap between upper and lower surrogates is defined as 

\begin{align}
    f_{\text{US}}(\tilde{N}) - f_{\text{LS}}(\tilde{N})  \nonumber \le \sum_k \left[ \ln\left(1 - \frac{1}{(\tilde{N}_k + 1)^2} e^{-c_k \tilde{N}_k}\right) - \ln\left(1 - e^{-c_k \tilde{N}_k}\right) \right] \nonumber
\end{align}

Using the first-order Taylor expansion $\ln(1 - x) \leq -x - \frac{x^2}{2}$ and discarding negative higher-order quadratic terms yields,
\begin{align}
 f_{\text{US}}(\tilde{N}) - f_{\text{LS}}(\tilde{N}) &\le \sum_{k=1}^L \left[ -\frac{1}{(\tilde{N}_k + 1)^2} e^{-c_k \tilde{N}_k} + e^{-c_k \tilde{N}_k} + \frac{1}{2} e^{-2c_k \tilde{N}_k} \right] \nonumber\\
&\le \sum_{k=1}^L e^{-c_k \tilde{N}_k} \left( 1 - \frac{1}{(\tilde{N}_k + 1)^2} + \frac{1}{2} e^{-c_k \tilde{N}_k} \right) . \nonumber
\end{align}

Since $\left[ 1 - \frac{1}{(\tilde{N}_k + 1)^2} + \frac{1}{2} e^{-c_k \tilde{N}_k} \right] < 2$, we can establish,

\begin{equation}
    f_{\text{US}}(\tilde{N}) - f_{\text{LS}}(\tilde{N})  \le 2\sum_{k=1}^L e^{-c_k \tilde{N}_k}. \label{eq:gap_bound_exp}
\end{equation}

From the optimality condition for the upper surrogate problem under multiplier $\lambda_{\text{US}}$, we have,
\begin{equation}
(\tilde{N}_k + 1)^2 e^{c_k \tilde{N}_k} - 1 = \frac{c_k}{\lambda_{\text{US}} w_k} + \frac{2}{(\tilde{N}_k+1) \lambda_{\text{US}} w_k}.
\label{US_optimal}
\end{equation}
Rearranging terms to isolate $e^{-c_k \tilde{N}_k}$ gives,
\begin{equation}
e^{-c_k \tilde{N}_k} = \frac{(\tilde{N}_k + 1)^2}{1 + \frac{c_k}{\lambda_{\text{US}} w_k}+\frac{2}{(\tilde{N}_k+1) \lambda_{\text{US}} w_k}} \le \frac{(\tilde{N}_k + 1)^2}{1 + \frac{c_k}{\lambda_{\text{US}} w_k}} \leq (\tilde{N}_k + 1)^2 \frac{\lambda_{\text{US}} w_k}{c_k} . \nonumber
\end{equation}

The last inequality follows from \( \frac{1}{1+u} \leq \frac{1}{u}\) for $u \geq 0$. Thus, we get
\begin{align}
f_{\text{US}}(\tilde{N}) - f_{\text{LS}}(\tilde{N})  &\leq 2 (\max_k \tilde{N}_k+1)^2 \cdot  \lambda_{\text{US}} \cdot \sum_{k=1}^L \frac{ w_k}{c_k}. 
\label{eq:gap_lambda_form}
\end{align}

Taking the logarithm of \eqref{US_optimal} yields $\tilde{N}_k + \frac{2}{c_k}\ln(\tilde{N}_k + 1) = \frac{1}{c_k} \ln\left(1 + \frac{c_k}{\lambda_{\text{US}} w_k}+\frac{2}{(\tilde{N}_k+1)\lambda_{US} w_k}\right)$. 

It follows from Lemma~\ref{lem:UBtildeNk} that

\begin{equation}
    \tilde{N}_k \le \frac{1}{c_k}\ln\left(1 + \frac{c_k}{\lambda_{\text{US}} w_k}\right).
    \label{N_k bound}
\end{equation}

From the upper bound on $\tilde{N}_k$ in \eqref{N_k bound}, we use the total budget constraint $\sum_{k=1}^L w_k \tilde{N}_k = W$, we obtain,
\begin{equation}
\sum_{k=1}^L \frac{w_k}{c_k} \ln\left(1 + \frac{c_k}{\lambda_{\text{US}} w_k}\right) \ge W.
\nonumber
\end{equation}

Define normalized probability weights $p_k$ (such that $\sum p_k = 1$),
\begin{equation}
    p_k = \frac{\frac{w_k}{c_k}}{\sum_u \frac{w_u}{c_u}}. \nonumber
\end{equation}

Dividing both sides of the constraint by $\sum \frac{w_k}{c_k}$ yields,
\begin{equation}
    \sum_k p_k \ln \left(1 + \frac{1}{p_k \sum \frac{w_k}{c_k}} \cdot \frac{1}{\lambda_{\text{US}}}\right) \ge \frac{W}{\sum \frac{w_k}{c_k}} \nonumber
\end{equation}

Because the natural logarithm function $\ln(x)$ is concave, we can use Jensen's Inequality to say that $\sum_k p_k \ln(X_k) \le \ln \left( \sum_k p_k X_k \right)$. Thus, the lower bound evaluates to

\begin{equation}
    \ln \left( \sum_{k=1}^L p_k \left( 1 + \frac{1}{p_k \sum \frac{w_k}{c_k}} \cdot \frac{1}{\lambda_{\text{US}}} \right)\right) \geq \frac{W}{\sum \frac{w_k}{c_k}}. \nonumber
\end{equation}

Evaluating the term inside the logarithm, we obtain,
\begin{align}
    \sum_{k=1}^L p_k \left( 1 + \frac{1}{p_k \sum \frac{w_k}{c_k}} \cdot \frac{1}{\lambda_{\text{US}}} \right) &= \sum_{k=1}^L p_k + \frac{1}{\sum \frac{w_k}{c_k}} \cdot \frac{1}{\lambda_{\text{US}}} \sum_{k=1}^L = 1 + \frac{L}{\sum \frac{w_k}{c_k}} \cdot \frac{1}{\lambda_{\text{US}}}  \nonumber
\end{align}

Therefore, we have a closed-from lower bound on $\frac{1}{\lambda_{\text{US}}}$ as,

\begin{align}
    1 + \frac{L}{\sum \frac{w_k}{c_k}} \cdot \frac{1}{\lambda_{\text{US}}} &\ge \exp\left( \frac{W}{\sum \frac{w_k}{c_k}} \right) \nonumber \\
    \frac{1}{\lambda_{\text{US}}} &\ge \left( e^{\frac{W}{\sum w_k/c_k}} - 1 \right) \left( \frac{\sum \frac{w_k}{c_k}}{L} \right).
    \label{eq:lambda_lower_US}
\end{align}

Since $W \gg \sum \frac{w_k}{c_k}$, we  can replace $(e^{\frac{W}{\sum w_k/c_k}} - 1)$ with $e^{\frac{W}{\sum w_k/c_k}}$.

Finally, plugging in \eqref{eq:lambda_lower_US}, the lower bound for $\frac{1}{\lambda_{\text{US}}}$ into the gap formula in \eqref{eq:gap_lambda_form} yields the closed-form upper bound,

\begin{align}
    f_{\text{US}}(\tilde{N}) - f_{\text{LS}}(\tilde{N}) &\le 2 (\max_k \tilde{N}_k+1)^2  \frac{L \cdot e^{\frac{-W}{\sum w_k/c_k}}}{\sum \frac{w_k}{c_k}} \sum \frac{w_k}{c_k} \nonumber \\
    &\leq  2 (\max_k \tilde{N}_k+1)^2 L \cdot e^{\frac{-W}{\sum w_k/c_k}} \nonumber .
\end{align}

We can analyze how the term \((\max_k \tilde{N}_k+1)^2\) affects the gap by considering the following cases.
\begin{itemize}
    \item Case 1 ($\max_k \frac{c_k}{\lambda_{US} w_k}<1$): From Lemma~\ref{lem:UBtildeNk}, $(\max_k \tilde{N}_k+1)^2$ is upper bounded by a constant.
    \item Case 2 ($\max_k \frac{c_k}{\lambda_{US} w_k}> 1$): Applying the inequality $\ln(1+x) \le \ln x + 1$ (valid for $x > 1$) to Lemma~\ref{lem:UBtildeNk} gives,
    \begin{equation}
    \tilde{N}_k \le \frac{1}{c_k} \ln\left(\frac{c_k}{\lambda_{\text{US}} w_k}\right) + \frac{1}{c_k} = \frac{1}{c_k} \ln\left(\frac{1}{\lambda_{\text{US}}}\right) + \frac{\ln c_k}{c_k} + \frac{1}{c_k}. \nonumber
    \end{equation}
    Since $w_k \ge 1$ (so $\ln w_k \ge 0$) and $\max_{c_k > 0} \frac{1 + \ln c_k}{c_k} = 1$, this simplifies to,
    \begin{equation}
    \tilde{N}_k \le \frac{1}{c_k} \ln\left(\frac{1}{\lambda_{\text{US}}}\right) + 1. \nonumber
    \end{equation}
\end{itemize}

In Case 2, from \eqref{eq:lambda_lower_US} we see that $\frac{1}{\lambda_{\text{US}}}$ scales as $\exp\left(\frac{W}{\sum_k w_k/c_k}\right)$. Thus, the term $\ln\left(\frac{1}{\lambda_{\text{US}}}\right)$ grows at most linearly with $\frac{W}{\sum_k w_k/c_k}$. Consequently, the term $(\max_k \tilde{N}_k+1)^2$ grows at most polynomially in $\frac{W}{\sum_k w_k/c_k}$, whereas the $\exp\left(-\frac{W}{\sum_k w_k/c_k}\right)$ term in \eqref{eq:gap_lambda_form} decays exponentially. 

In the high-budget regime $W \gg \sum_{k=1}^L \frac{w_k}{c_k}$ (which is satisfied under the scaling condition $W \ge \ln(L) \sum_{k=1}^L \frac{w_k}{c_k}$ for large $L$), the exponential decay strictly dominates any polynomial growth. Thus, the bound on the gap in \eqref{eq:gap_lambda_form} still behaves as \( \lambda_{\text{US}} \cdot \sum_{k=1}^L \frac{ w_k}{c_k}\).

Hence, the upper bound on the surrogate gap becomes,
\begin{equation}
f_{\text{US}}(\tilde{N}) - f_{\text{LS}}(\tilde{N}) \le O\left(L \exp\left(-\frac{W}{\sum_{k=1}^L \frac{w_k}{c_k}}\right)\right). \nonumber
\end{equation}

The minimum budget scaling rule needed so that the gap is vanishing with $L$ (and $W$) is

\begin{equation}
\frac{W}{\sum_{k=1}^L \frac{w_k}{c_k}} = \ln(L) + \omega(1). \nonumber
\end{equation}

This gives us the gap, 

\begin{equation}
f_{\text{US}}(\tilde{N}) - f_{\text{LS}}(\tilde{N})  \le O\left(L \cdot \exp\left( - \big( \ln(L) + \omega(1) \big) \right) \right)= O\left(2L \cdot \frac{1}{L} \cdot e^{-\omega(1)}\right) = o(1). \nonumber
\end{equation}
Thus, a budget scaling exceeding logarithmic growth in $L$ (relative to $\sum_{k=1}^L w_k/c_k$) is sufficient for the accuracy achieved by the optimizer of the lower surrogate to be within $1-o(1)$ factor of the optimum accuracy.

\end{proof}

\begin{lemma}
\label{lem:UBtildeNk}
Let $N_{\max} = \frac{1}{c_k} \ln\left(1 + \frac{c_k}{\lambda_{\text{US}} w_k}\right)$. The upper surrogate allocation $\tilde{N}_k$ satisfies $\tilde{N}_k \le N_{\max}$.
\end{lemma}
\begin{proof}

Define a function
\begin{equation}
    f(N) = N + \frac{2}{c_k}\ln(N + 1) - \frac{1}{c_k} \ln\left(1 + \frac{c_k}{\lambda_{\text{US}} w_k} + \frac{2}{(N+1)\lambda_{\text{US}} w_k}\right). \nonumber
\end{equation}
Observe that for $N \ge 0$, both $N$ and $\frac{2}{c_k}\ln(N+1)$ are strictly increasing, while $\frac{2}{(N+1)\lambda_{\text{US}} w_k}$ is strictly decreasing, making $-\frac{1}{c_k}\ln\left(1 + \frac{c_k}{\lambda_{\text{US}} w_k} + \frac{2}{(N+1)\lambda_{\text{US}} w_k}\right)$ strictly increasing as well. 

Consequently, $f(N)$ is strictly increasing on $[0, \infty)$. Since $f(\tilde{N}_k) = 0$, proving $\tilde{N}_k \le N_{\max}$ is equivalent to proving $f(N_{\max}) \ge 0$.

From the definition of $N_{\max}$, we have
\begin{equation}
    e^{c_k N_{\max}} = 1 + \frac{c_k}{\lambda_{\text{US}} w_k} \implies \frac{1}{\lambda_{\text{US}} w_k} = \frac{e^{c_k N_{\max}} - 1}{c_k}. \nonumber
\end{equation}
Evaluating $f(N)$ at $N = N_{\max}$ and combining the logarithmic terms gives
\begin{equation}
    f(N_{\max}) = \frac{1}{c_k}\ln\left(e^{c_k N_{\max}} (N_{\max} + 1)^2\right) - \frac{1}{c_k}\ln\left(e^{c_k N_{\max}} + \frac{2}{(N_{\max}+1)\lambda_{\text{US}} w_k}\right). \nonumber
\end{equation}
Since $\frac{1}{c_k}\ln(\cdot)$ is strictly increasing, $f(N_{\max}) > 0$ holds if and only if
\begin{equation}
    e^{c_k N_{\max}} (N_{\max} + 1)^2 > e^{c_k N_{\max}} + \frac{2}{(N_{\max}+1)\lambda_{\text{US}} w_k}. \nonumber
\end{equation}
Subtracting $e^{c_k N_{\max}}$ from both sides, noting that $(N_{\max} + 1)^2 - 1 = N_{\max}(N_{\max} + 2)$, and substituting $\frac{1}{\lambda_{\text{US}} w_k} = \frac{e^{c_k N_{\max}} - 1}{c_k}$ yields
\begin{equation}
    e^{c_k N_{\max}} N_{\max}(N_{\max} + 2) > \frac{2}{N_{\max}+1} \cdot \frac{e^{c_k N_{\max}} - 1}{c_k}. \nonumber
\end{equation}
Letting $z = c_k N_{\max} > 0$ and dividing both sides by $N_{\max} > 0$, this simplifies to
\begin{equation}
    e^z (N_{\max} + 2) > \frac{2}{N_{\max} + 1} \cdot \frac{e^z - 1}{z}. \nonumber
\end{equation}
This inequality holds strictly for all $N_{\max} > 0$ and $z > 0$ because:
\begin{enumerate}
    \item $N_{\max} + 2 > 2 > \frac{2}{N_{\max} + 1}$ (since $N_{\max} + 1 > 1$), and
    \item $e^z > \frac{e^z - 1}{z}$ (since $e^z - 1 = \int_0^z e^t \, \mathrm{d}t < z e^z$ for $z > 0$).
\end{enumerate}
Therefore, $f(N_{\max}) > 0 = f(\tilde{N}_k)$. By the strict monotonicity of $f(N)$, it follows that
\begin{equation}
    \tilde{N}_k < \frac{1}{c_k}\ln\left(1 + \frac{c_k}{\lambda_{\text{US}} w_k}\right), \nonumber
\end{equation}
which establishes the desired upper bound.
\end{proof}

\begin{lemma}
\label{lem:ckwk_condition}
The condition $\frac{c_k}{\lambda_{\text{LS}} w_k} \gg 1$  and  \(\lambda_{\text{LS}}  = \exp \left( -\frac{W + \sum_{k=1}^L \frac{w_k}{c_k} \ln\left(\frac{w_k}{c_k}\right)}{\sum_{k=1}^L \frac{w_k}{c_k}} \right)
\) hold if the total budget satisfies $W=\sum_{k=1}^L\frac{w_k}{c_k} (\ln L + \omega(1))$.
\end{lemma}

\begin{proof}
Let $a_k = \frac{w_k}{c_k}$, so
the $\frac{c_k}{\lambda_{\text{LS}} w_k} \gg 1$ would be equivalent to proving $a_k \ll \frac{1}{\lambda_{\text{LS}}}$. A sufficient condition for proving $a_k \ll \frac{1}{\lambda_{\text{LS}}}$ is $\ln a_k + \omega(1) \le \ln \frac{1}{\lambda_{\text{LS}}}$.

Substituting the derived expression for $\frac{1}{\lambda_{\text{LS}}}$ from \eqref{eq:lambda_exp}, we have, 
\begin{equation}
    \ln \left( \frac{1}{\lambda_{\text{LS}}}\right)  =   \frac{W + \sum_{k=1}^L \frac{w_k}{c_k} \ln\left(\frac{w_k}{c_k}\right)}{\sum_{k=1}^L \frac{w_k}{c_k}}. \nonumber
\end{equation}

Thus, for proving the condition for $a_k$ it is sufficient to prove

\begin{equation}
\ln a_k + \omega(1) \le \frac{W + \sum_{k=1}^L a_k \ln a_k}{\sum_{k=1}^La_k}. \nonumber
% \label{eq:to_prove}
\end{equation}

Let us define probability weights $a_k/A$ for $A = \sum_{k=1}^La_k$ such that $\sum_{k=1}^L \frac{a_k}{A} = 1$. 

we expand the term \(\frac{\sum_{k=1}^L a_k \ln a_k}{A}\) as 
\begin{equation}
    \frac{1}{A}\sum_{k=1}^L a_k \ln a_k = \ln A +  \sum_{k=1}^L \frac{a_k}{A} \ln \left( \frac{a_k}{A} \right) . \nonumber
\end{equation}

This finally gives us the simplified expression for the RHS term as,
\begin{equation}
 \frac{W}{A} +  \ln A +  \sum_{k=1}^L \frac{a_k}{A} \ln \left( \frac{a_k}{A} \right). \nonumber
\end{equation}

Now, the term  \(\sum_{k=1}^L \frac{a_k}{A} \ln \left( \frac{a_k}{A} \right) = -H\left(\{\frac{a_k}{A} \}\right)\) is the Shannon entropy of the probability \(\frac{a_k}{A}\). Thus, the expression becomes,

\begin{equation}
\frac{W}{A} +  \ln A - H\left(\{\frac{a_k}{A} \}\right) \nonumber
\end{equation}

By the property of Shannon entropy we have, $H\left(\{\frac{a_k}{A}\}\right) \le \ln L$.

Since $\ln A=\ln \sum_k a_k \ge \ln a_k$ and in Theorem~\ref{thm:asymptotic_optimality}, $W = A (\ln L+\omega(1))$, we have $\ln a_k + \omega(1) \le \frac{W + \sum_{k=1}^L a_k \ln a_k}{\sum_{k=1}^La_k}$.

\end{proof}

\end{document}